\documentclass[preprint,5p,times,twocolumn,square,sort&compress,numbers]{elsarticle}

\usepackage{subcaption}
\usepackage{mathtools}
\usepackage[colorlinks]{hyperref}
\usepackage[super]{nth}
\usepackage[export]{adjustbox}
\usepackage{amsmath,amssymb}
\usepackage{stfloats}
\usepackage{overpic}

\journal{Ultramicroscopy}

\newif\ifNOSUP \NOSUPtrue

\begin{document}

\begin{frontmatter}

\title{Large-Angle Convergent-Beam Electron Diffraction Patterns via Conditional Generative Adversarial Networks
}

\author[addwarwick,addoxford]{Joseph J Webb\corref{cor}} 
\ead{webb@maths.ox.ac.uk}
\author[addwarwick]{Richard Beanland}
\ead{r.beanland@warwick.ac.uk}
\author[addwarwick]{Rudolf A Römer}
\ead{r.roemer@warwick.ac.uk}

\address[addwarwick]{Department of Physics, University of Warwick, Gibbet Hill Road, Coventry, CV4 7AL, United Kingdom}
\address[addoxford]{Mathematical Institute, University of Oxford, Andrew Wiles Building, Radcliffe Observatory Quarter, Woodstock Road, Oxford, OX2 6GG, United Kingdom}

\cortext[cor]{Corresponding author}


\begin{abstract}
We show how generative machine learning can be used for the rapid computation of strongly dynamical electron diffraction directly from crystal structures, specifically in large-angle convergent-beam electron diffraction (LACBED) patterns. We find that a conditional generative adversarial network can learn the connection between the projected potential from a cubic crystal's unit cell and the corresponding LACBED pattern. Our model can generate diffraction patterns on a GPU many orders of magnitude faster than existing direct simulation methods.
Furthermore, our approach can accurately retrieve the projected potential from diffraction patterns, opening a new approach for the inverse problem of determining crystal structure.
\end{abstract}


\end{frontmatter}

\section{Introduction} \label{introduction}

Convergent-beam electron diffraction (CBED) \cite{Spence1992ElectronMicrodiffraction,Tanaka1994Convergent-beamDiffraction} is a transmission electron microscopy (TEM) technique with unparalleled sensitivity \cite{Beanland2013DigitalPicture}. Its origins date back nearly $100$ years to pioneering work by \citet{Kossel1939ElektroneninterferenzenBundel} and its modern applications include crystal symmetry classification \cite{Buxton1976ThePatterns, Tanaka1983Point-groupDiffraction, Tanaka1983Space-groupDiffraction}, lattice parameter determination \cite{Saunders1999QuantitativeMeasurements, Zuo1998AMatching, Kaiser1999ApplicationSubstrates}, strain \& defect analysis \cite{Armigliato2003ApplicationDevices, Kramer2000AnalysisCBED, Cherns1989ConvergentMultilayers, Morniroli1996AnalysisDiffraction}, and more \cite{Midgley1996QuantitativeBonds}. 
However, CBED sees the majority of its use in symmetry determination \cite{Buxton1976ThePatterns} and charge density refinement \cite{Zuo1999DirectCu2O} and is still lacking in popularity when compared to the more established structure solution and refinement methods of X-ray and neutron diffraction  \cite{Beanland2021RefinementDiffraction, Hubert2019StructurePatterns}.
Collecting the necessary amount of high-quality diffraction data from a TEM, to construct a LACBED image, is one of the inherent challenges of the method. Here, modern computer-controlled TEM offers a clear advantage and can make the task near automatic \cite{Beanland2013DigitalPicture,Busch,koch2011,koch2023}. 

A perhaps even more constraining challenge lies in the fact that the complexity introduced by multiple scattering of electrons as they propagate through the specimen \cite{Spence1992ElectronMicrodiffraction} requires sophisticated modelling techniques to compare with TEM results.
To make CBED quantitative, there have been two major computational methods developed: (i) the Bloch-wave method \cite{Zuo1995OnMethod, Spence1992ElectronMicrodiffraction, 2018ElectronMicroscopy, Hubert2019StructurePatterns}, and (ii) Multislice \cite{Cowley1957TheApproach,  VanDyck1984TheMicroscopy, Chuvilin2005OnSimulation, Kaiser2006ProspectsCalculation, Kirkland2010AdvancedMicroscopy, 2018ElectronMicroscopy}. Whilst both have seen success in accurately generating CBED patterns, they even today remain computationally resource- and time-intensive, often well beyond what a standard desktop computer can provide \cite{BeanlandFelix:Software}.

In this work, we show that recent advances in machine learning offer an exciting way to circumvent this obstacle. We use a \emph{generative} deep learning architecture to readily predict bright field LACBED patterns with high precision.
The application of machine learning to electron microscopy has blossomed in the last decade \cite{Ede2021AdvancesLearning}, in line with the uptake of machine learning across nearly the full breadth of the natural sciences \cite{Carleo2016,Mehta2018}. For example, strategies of machine learning have been used to reduce the data flow in single-molecule data classification \cite{Matinyan2023MachineData}, convolutional neural nets were shown to help with phase reconstruction for CBED-based scanning TEM \cite{Friedrich2023PhaseLearning} while molecular structure imaging was found to benefit from such CNNs as well \cite{Liu2021MachineStructures}. At the core of the deep learning methods employed in these works lies the astonishing progress in the last decades in so-called \emph{supervised} learning techniques now routinely employed across search engines and computer vision applications \cite{2023Kaggle:Science,Wichert2021MachineLearning}.

\begin{figure}[tb]
    \centering
    Normalised potential\\
    \includegraphics[scale=1]{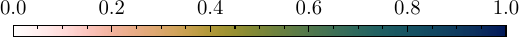}\\[3pt]
    \subfloat[\centering Li$_{21.2}$Ge$_5$ ]{\includegraphics[width=0.3\columnwidth]{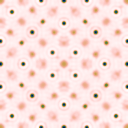}} \hspace{3pt}
    \subfloat[\centering ZnC$_4$O$_4$(H$_2$O)$_2$]
    {\includegraphics[width=0.3\columnwidth]{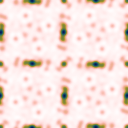}} \hspace{3pt}
    \subfloat[\centering Cd$_25$Eu$_4$]{\includegraphics[width=0.3\columnwidth]{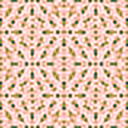}}

\vspace{3pt}
Mean prediction \\[1pt]
\includegraphics[width=0.3\columnwidth]{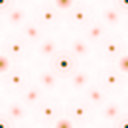}\hspace{3pt}        \includegraphics[width=0.3\columnwidth]{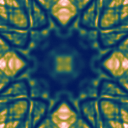}\hspace{3pt}        \includegraphics[width=0.3\columnwidth]{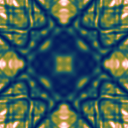}\hspace{3pt}\\[6pt]
Median prediction \\[1pt]
\includegraphics[width=0.3\columnwidth]{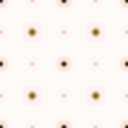}\hspace{3pt}        \includegraphics[width=0.3\columnwidth]{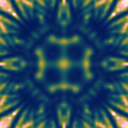}\hspace{3pt}        \includegraphics[width=0.3\columnwidth]{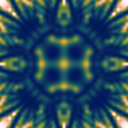}\hspace{3pt}\\[6pt]
    
    \centering
    Intensity\\
    \includegraphics[scale=1]{figs/batloww_r_colorbar.pdf}\\[2pt]

\caption{Top: Three examples of a $128\times 128$ normalised $[001]$ projected potential $\rho(\mathbf{r})$, following Eq.\ \eqref{eq:projpot}. 
(a) Li$_{21.2}$Ge$_5$, ICSD 93421, $F\bar{4}3m$, $a=18.756\AA$; 
(b) ZnC$_4$O$_4$(H$_2$O)$_2$, ICSD 95687, $Pn\bar{3}$, $a=16.256\AA$; (c) Cd$_25$Eu$_4$, ICSD 252137, $Fd\bar{3}$, $a=31.872\AA$.  
Bottom: bright field LACBED patterns for mean cGAN performance $R=0.966$ (CeMnNi$_4$, ICSD 262460, top) and median cGAN performance $R=0.993$ (SrFeO$_3$, ICSD {154938}, bottom).  
Left: [001] projected potential of a unit cell; centre: cGAN LACBED pattern; right: {\textsc{Felix}} Bloch-wave simulation (ground truth). All images are $128\times 128$ pixels and have normalised intensities.}
\label{fig:projpot-ML-Felix}
\end{figure}

The power of \emph{generative} machine learning has not yet been harnessed to the same extent, neither in physics, in general, \cite{Mehta2018} nor in TEM, in particular. This is partly because it is still a relatively novel machine learning strategy \cite{Wichert2021MachineLearning}. The generative method can \emph{create} novel predictions which do not appear in any of the provided data, unlike supervised methods that generally interpolate between training data points.  For example, in computer vision, generative networks construct previously non-existent high-resolution images, conditional on information from other images \cite{Isola2017Image-to-imageNetworks, Wang2018High-ResolutionGANs}.
Here, we use this strategy to create LACBED patterns by providing the projected electron density as input (Figure \ref{fig:projpot-ML-Felix}).  We ignore contributions from higher-order Laue zones (HOLZ), allowing only the projected potential to be needed as input for a given material.
Each reference LACBED pattern, taken to be the desired output for the supervised ML task, required about $200$ seconds to be constructed by the Bloch-wave method on 32 cores of a high-performance compute cluster, while our cGAN LACBED images arrived within $20$ milliseconds on a modern, i.e.\ GPU-supported, desktop. 

In the remainder of the manuscript, we describe how our approach uses a conditional generative adversarial network (cGAN) to produce the simulations shown in Fig. \ref{fig:projpot-ML-Felix}.  
As an initial exploration of the approach, we simplify the problem to make it readily amenable to a machine-learning strategy. Thus, we choose both input and output images to have dimensions of $128\times128$ pixels.  This size is sufficient for many computer-vision-based machine learning tasks \cite{Wichert2021MachineLearning}, whilst remaining small enough to allow generation of results on a large scale.
To be readily consistent with this geometry we encode the crystal structure in the form of a normalised projected potential of the unit cell of an inorganic cubic crystal (see section~\ref{sec:icsd}) aligned to the [001] zone axis, and assemble this information as input data for subsequent machine learning tasks as discussed in section \ref{sec:input_data}.  
The desired output is the direct beam [001] LACBED pattern at a single crystal thickness.  The angular range of the LACBED pattern is scaled in inverse proportion to lattice parameter, ensuring that it contains strong dynamical diffraction effects across its full area and does not contain large blank areas. 

The high symmetry of the input data results in a high symmetry of the simulated patterns.  This symmetry was not constrained in the cGAN calculations and provides an additional check on the output.
%
%
%
%
The details of the machine learning methodology and cGAN architecture are covered in Section \ref{sec:machine_learning_methodology}. 
In Section \ref{sec:results}, we present and discuss our results, showing that using this approach one can generate accurate diffraction patterns many orders of magnitude faster than current methods. We also find that we can solve the inverse problem, namely, reconstruct the projected potential from diffraction patterns. 
We discuss the future of our approach and improvements that can be made in Section \ref{sec:discussion_and_conclusions}. 

\section{Data}
\label{sec:data}

\subsection{Selection of training data}
\label{sec:icsd}
\label{sec:data-training}

To generate LACBED patterns via machine learning, we require a large body of data in which they are paired with crystal structure. Previous machine learning in computer vision \cite{Wichert2021MachineLearning,2023Kaggle:Science} and related applications \cite{Ohtsuki2019,Ohtsuki2019a,Acevedo2021,Bayo2022,Bayo2023} suggests that often more than $10,000$ such training pairs are needed. It is infeasible to use experimental data on this scale and we therefore use simulations with structures taken from the Inorganic Crystal Structure Database (ICSD) \cite{IgorLevin2018NISTICSD}, which contains more than $240,000$ structures.  We chose cubic crystal structures with a publication year of 2000 or later, giving $21,601$ Crystallographic Information Files (CIFs) \cite{Hall2006SpecificationCIF}. Of these, we took $14,270$ structures with unique chemical formulae as our ML data set.  Direct training with textual data such as a CIFs is still a major challenge for machine learning \cite{Halevy2009TheData,Spasic2020ClinicalReview}. Thus, for each structure we (i) computed, via Bloch-wave code \textsc{Felix}, a bright field LACBED image as ground truth and (ii) calculated the corresponding normalised projected electronic potential to serve as cGAN input data.  The input data spanned all $36$ cubic space groups, giving $6$ different plane group symmetries in the two-dimensional projected potential.

    
    \label{fig:projected_potential}

\subsection{Input data}
\label{sec:input_data}

 We calculate the $[001]$ projected potential $\rho$ in a $128\times128$ image of the unit cell, as shown in Fig. \ref{fig:projpot-ML-Felix}, using a Fourier series of structure factors $F(\mathbf{g})$ calculated in \textsc{Felix} for each crystal, i.e.
\begin{equation}
 \rho (\mathbf{r}) = \dfrac{1}{V} \sum_{\mathbf{g}} F (\mathbf{g}) \cdot \exp[-2\pi i \mathbf{g} \cdot \mathbf{r}],  
 \label{eq:projpot}
\end{equation}
where the series is truncated after 2500 reciprocal lattice vectors $\mathbf{g}$.  We normalise the resulting potential to the integer range 0-255. 
In Fig.\ \ref{fig:projpot-ML-Felix} we show three examples of $\rho (\mathbf{r})$ constructed in this way.
By requiring all machine learning model inputs to have the same dimensions some information regarding the size of the unit cell is lost (discussed further in Section \hyperref[sec:discussion_and_conclusions]{5}).

\subsection{Simulation Software}
\label{sec:felix}

\textsc{Felix} is an open-source implementation of the Bloch-wave method for generating LACBED images \cite{BeanlandFelix:Software, 2018ElectronMicroscopy,Zuo1995OnMethod}.  The software takes as input a CIF, microscope and crystal settings, and the number of  beams to be included in the Bloch wave calculation.  It has been shown to provide atomic coordinate refinements with picometer accuracy \cite{Beanland2021RefinementDiffraction,Hubert2019StructurePatterns}, and can accurately simulate LACBED patterns when compared to experimental data 
\cite{Beanland2013DigitalPicture}.

In this investigation, for simplicity we only consider the direct beam, using \textsc{Felix} to produce bright field LACBED patterns at specimen thicknesses of 50, 100, 150 and 200 nm.  
The number of beams in the calculation was scaled in proportion with the unit cell dimension $a$, giving a roughly constant resolution limit in reciprocal space of $3.5 \AA^{-1}$, except for the largest unit cells (with $a \geq 14 \AA$) where the number of beams in the Bloch wave calculation was limited to $2500$ (amounting to $\sim 13\%$ of our data files). 
The half-convergence angle $\alpha$ was scaled in inverse proportion to the unit cell, $\alpha=0.21/a$, placing the limit of the bright field LACBED pattern roughly at the 10~0~0 Bragg condition. Absorption was neglected. 
Calculation times varied from 100 to 1500 seconds, depending on, amongst others, the number of beams and the number of atoms in the unit cell. 
All simulation parameters are provided in the code accompanying the present work \cite{Webb2024GitHubWephy/ai-diffraction}. Including scheduling and computation, generating the ground-truth LACBED dataset at four specimen thicknesses required several weeks using a high-performance cluster, providing a dataset of $4 \times 14,270$ image pairs \cite{Webb2023FelixPatterns}.

\section{Machine Learning Methodology}
\label{sec:machine_learning_methodology}

\begin{figure*}[tb]
    \centering
    \includegraphics[width=0.95\textwidth]{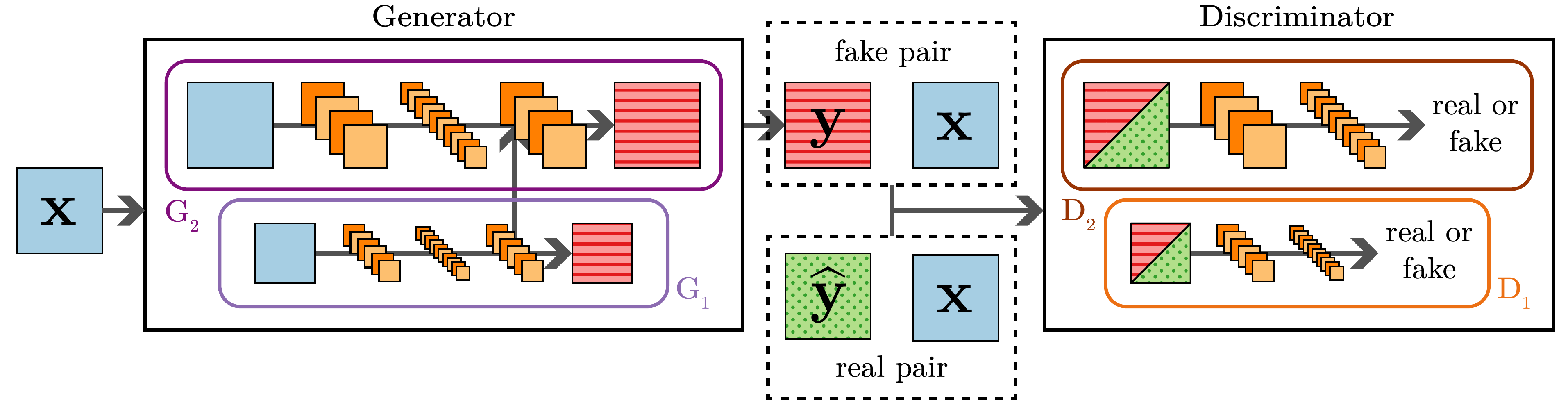}
    \caption{Schematic of our cGAN architecture with $\mathbf{x}$ the input (blue squares), $\mathbf{y}$ the generated prediction (red dashed square) and $\hat{\mathbf{y}}$ the ground truth (green dotted square). 
    Generators are labelled G$_1$, G$_2$ and discriminators are D$_1$, D$_2$. 
    The orange squares indicate convolution layers with ReLU activations and their different sizes schematically denote the downsampling according to chosen kernel sizes.
    We note that $\mathbf{x}$ and $\hat{\mathbf{y}}$ correspond to $\rho$ and \textsc{Felix}-simulated LACBED images for CBED prediction as in section \ref{sec:diffraction_patterns}, 
    or, conversely, to \textsc{Felix}-simulated CBED images and their generated 
    $\rho$ images for $\rho$ prediction as in section \ref{sec:electron_potential}.}
    \label{fig:cGAN-algo-structure}
\end{figure*}

\subsection{Design considerations}

Currently, the most popular image-to-image translation architectures are generative adversarial networks (GANs) \cite{IanGoodfellow2016DeepLearning, Goodfellow2014GenerativeNets} and variational autoencoders (VAEs) \cite{IanGoodfellow2016DeepLearning, Kingma2013Auto-EncodingBayes, Goodfellow2014GenerativeNets}. Autoencoders focus on learning two functions: one to encode input data into a latent vector, and one to decode this to output data, traditionally to recreate the input data. However, instead of simply recreating input data, it can be used as a generative model, under the assumption that the generated data and input data share structural information which manifests in the latent vector.
Nevertheless, VAEs can suffer from the blurring of high-fidelity output \cite{Pathak2016ContextInpaintingb}, which would be problematic for quantitative electron diffraction. This issue does not occur for GANs \cite{Isola2017Image-to-imageNetworks}. We thus use a conditional GAN (cGAN) \cite{Mirza2014ConditionalNets} in which a mapping from an image $\mathbf{x}$ and random noise vector $\mathbf{z}$ to another image $\mathbf{y}$, $G: \{\mathbf{x}, \mathbf{z}\} \to \mathbf{y}$, is learned. Here, $G$ is called the \emph{generator}. GANs also feature another object called the \emph{discriminator}, $D$, which is trained to discern between `real' images from the dataset, and `fake' images from the generator. 
Whilst VAEs require a predefined loss function, the parameters in GANs are instead optimised via competition between the discriminator and the generator (Fig.\ \ref{fig:cGAN-algo-structure}). Specifically, this is achieved with 
$
\min_{G} \max_{D} \mathcal{L}_{\text{GAN}}(G, D)$ (a minimax game \cite{StuartRussellArtificialApproach}),
where 
\begin{equation}
\mathcal{L}_{\text{GAN}}(G,D) =
\mathbb{E}_{(\mathbf{x},\hat{\mathbf{y}})}
\left[\log D(\mathbf{x}, \hat{\mathbf{y}})\right]
+
\mathbb{E}_{\mathbf{x}}
\left[
\log\left(1-D(\mathbf{x}, G(\mathbf{x})\right)
\right]
\end{equation}
is the \emph{objective function} and $\hat{\mathbf{y}}$ is the ground truth.
The process involves taking alternative steps between optimising $G$ and optimising $D$. This solves the challenge of having to find an optimal loss function for comparing dynamical diffraction patterns, which is not \emph{a priori} clear.

\subsection{Architecture and Implementation}

We use the \textsc{pix2pix} architecture, specifically that in \citet{Wang2018High-ResolutionGANs} which develops what is known as \textsc{pix2pixHD}, building on work by \citet{Isola2017Image-to-imageNetworks} and \citet{Radford2015UnsupervisedNetworks}. 
Briefly, this architecture uses two generators, operating in tandem at reduced and full pixel resolution to generate predictions, while two discriminators judge their real/fake predictions at two image resolution levels.
We train up to a maximum of $\varepsilon_\text{max} = 100$ epochs with a learning rate of $\ell=2\times 10^{-4}$ using the \textsc{Adam} optimizer \cite{Wang2018High-ResolutionGANs}. Our models are not pre-trained.\footnote{This is by design since neural nets with weights pre-trained on LACBED images do not exist and the images are very different from more common image tasks in computer vision. In particular, the symmetry of the CBED images will play a major role so that networks pre-trained with different symmetries or an absence of such are likely detrimental to performance.}

\begin{figure}[tb]
    \centering
    \includegraphics[scale=1]{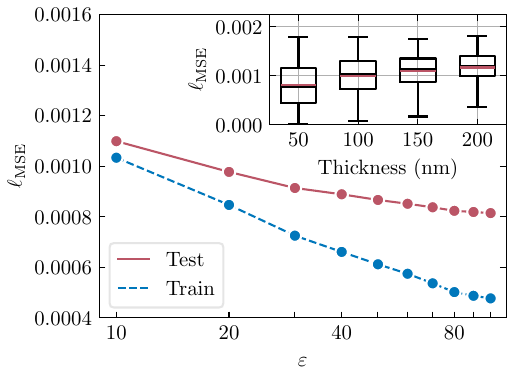}
    \caption{MSE loss $\ell_\text{MSE}$ for all data as a function of epoch $\varepsilon$ for a specimen thickness of $500$ nm. Blue (Red) circles show average loss over $10$ independent training (test) runs. Errors are smaller than the symbols.  Note the logarithmic scale for $\varepsilon$.
    Inset: mean (red lines) and median (black lines) of $\ell_\text{MSE}$ at $\varepsilon=100$ for the test data in box-plot representation for four specimen thicknesses. 
    The vertical size of the box gives the $25$th and $75$th percentiles and the error bars denote the error of the mean.
    }
    \label{fig:losses}
\end{figure}

As discussed above, there is no concrete loss function when \emph{training} GANs, that is, GANs do not feature a function which produces a loss from the pair $(\mathbf{y}, \mathbf{\hat{y}})$.
%
Thus, when quantifying predictions \emph{after training}, we are free to choose convenient loss functions for consistent and reproducable comparison between generated images.
Let $i,j= 1, \ldots, n$ denote pixel indices in each $n \times n$ image $\mathbf{y}=\{y_{ij}\}$, and \textsc{Felix} simulated ground truth $\hat{\mathbf{y}}=\{\hat{y}_{ij}\}$. We employ the per-pixel MSE loss function, 
%
\begin{equation}
    \ell_\text{MSE}(\mathbf{y}, \hat{\mathbf{y}}) = \frac{1}{n^2} \sum_{i,j}^{n} (y_{ij} - \hat{y}_{ij})^2,
    \label{eq:mse}
\end{equation}
to evaluate training and validation convergence of our cGAN (Fig.~\ref{fig:losses}).
%
In addition, we use a modified zero-mean normalised cross-correlation index for pixel intensities \cite{Beanland2013DigitalPicture},
\begin{equation}
    R(\mathbf{y}, \hat{\mathbf{y}}) = \frac{1}{2} + \frac{1}{2 n^2}\sum_{i,j}^{n} 
    \frac{y_{ij}-\langle \mathbf{y} \rangle}{\sigma(\mathbf{y})} \cdot
    \frac{\hat{y}_{ij}-\langle \hat{\mathbf{y}} \rangle}{\sigma(\hat{\mathbf{y}})} ,
    \label{eq:zmcc}
\end{equation}
where $\langle \mathbf{y} \rangle$ and $\sigma(\mathbf{y})$ denote the mean and standard deviation of pixel intensities in $\mathbf{y}$, and similarly for $\hat{\mathbf{y}}$. $R=1$ corresponds to a perfect fit, while $R=0$ is perfectly anti-correlated and $R=0.5$ corresponds to two images with uncorrelated intensities.
Thirdly, we measure \emph{local} differences between $\mathbf{y}$ and $\hat{\mathbf{y}}$ using the pixel-resolved squared error
\begin{equation}
    p(i,j) = \left(y_{ij} - \hat{y}_{ij}\right)^2,
    \label{eq:pse}
\end{equation}
 When viewed as an image, $p(i,j)$ allows areas of good and poor performance to be identified. 

Standard errors are given as an average performance using ten-fold cross-validation, in which the data is partitioned into ten random subsets of equal size. Ten models are then trained, each with nine subsets as training data and each judged by its performance on the remaining, test, subset. This method allows for every input to be tested with a model that has not been trained on it, and allows for a much larger set of test cases, providing a more accurate assessment for a relatively small dataset.  All results presented here (e.g. Figs.\ref{fig:projpot-ML-Felix}, \ref{fig:pred-lacbed}) are taken from the unseen test set at $\varepsilon=100$. Training was performed on a on an NVIDIA RTX 2070 SUPER GPU with a batch size of $16$.
%

\section{Results}
\label{sec:results}


\subsection{Simulation of Bright Field Diffraction Patterns}
\label{sec:diffraction_patterns}

Fig.~\ref{fig:losses} shows the improvement in training and test losses with increasing $\varepsilon$.  The difference between seen (Train) and unseen (Test) calculations increases as training proceeds, until at $\varepsilon=100$ we have $\ell_\text{MSE, test} = 8.15(4)\times 10^{-4}$ and $\ell_\text{MSE, train} = 4.773(9)\times 10^{-4}$ for a specimen thicknesses of $50$ nm. While these values indicate a close agreement between the ML and Bloch-wave simulations the improvement in $\ell_\text{MSE}$ reduces exponentially with $\varepsilon$ as is apparent in Fig.~\ref{fig:losses} from the constant gradient when plotted with log-linear axes, and has essentially come to a halt for the test set at $\varepsilon=100$.  
We do not go beyond this point as further training is likely to lead to severe over-fitting.

\begin{figure*}[tb]
    \centering
    \begin{overpic}[width=2.0\columnwidth,grid=false]{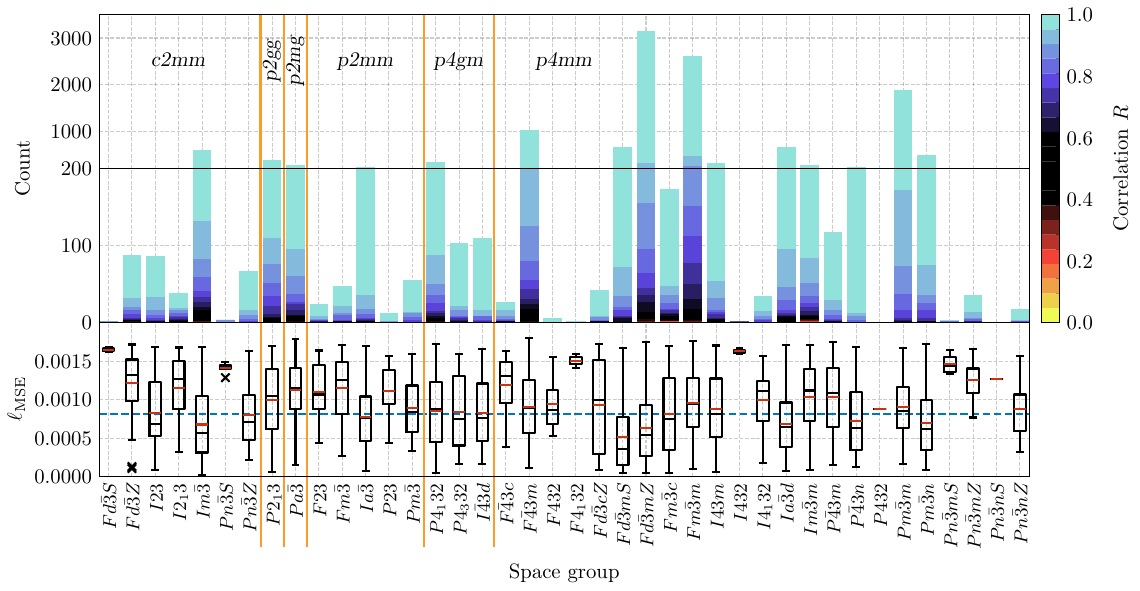}
    \put(-1.5, 49){(a)} 
    \put(-1.5, 22){(b)} 
    \end{overpic}
    \caption{(a) The distribution of crystal structure data entries in each of the $36$ ($+4$ from alternative origin choices) cubic space groups as obtained from the ICSD. 
    The colour scale denotes the cross-correlation index $R$, see \eqref{eq:zmcc}, obtained for each structure in the indicated space group using the trained cGAN, sorted from overall lowest (yellow) to overall highest (cyan) correlation. 
    The vertical scale above $200$ has been compressed for clarity.
    (b) Standard box-plot (cp.\ Fig.\ \ref{fig:losses}) of the median (black line inside a box), mean (red line inside a box), and $25$th and $75$th percentiles, denoting the vertical size of the box, each presenting their respective loss value $\ell_{\text{MSE}}$ for unseen crystal structures. 
    Outliers are denoted by crosses which lie more than $1.5$ $\times$ the inter-quartile range from the box. The horizontal dashed line highlights the overall mean loss $\ell_{\text{MSE}} = 0.00815(4)$ across unseen crystal structures.} 
    \label{fig:data-loss}
\end{figure*}
The behaviour of the cGAN simulation for the different space groups in the input data is shown in Fig.~\ref{fig:data-loss} in terms of (a) correlation coefficient $R$ and (b) $\ell_\text{MSE}$. The average $\ell_\text{MSE}$ is below 0.001 for the majority of space groups, indicating good overall performance.  Due to the distribution of experimental ICSD data across the different space groups, the vast majority of the input data has the plane group $\textsl{p4mm}$, but all plane groups contain more than $200$ ICSD entries.  
However, space groups with only a few examples, such as $F4_{1}32$ and $I432$, have noticeably larger $\ell_\text{MSE} \approx 0.0015$, and this occurs even though they have the same plane group as others that are much better.  It is known that imbalanced data can affect the predictive strength of adversarial networks \cite{Saini2023TacklingReview}, but this result indicates that there is no influence of (projected) input symmetry on the quality of the result.  The factor determining the quality of cGAN simulations is probably due to differences in the position of atoms in the unit cell, which will vary significantly for different space groups.  This suggestion is supported by the large discrepancy and poor performance for the small number of space groups with alternative ($S$) origin choices rather than centred ($Z$), such as $Fd\bar{3}$, $Pn\bar{3}$ and $Pn\bar{3}n$.  The choice of origin makes no physical difference to a Bloch wave simulation, but changes the position of atoms in the projected potential used as cGAN input.  This change clearly has an unwanted impact on the output of our calculation.

\begin{figure}[tb]
    \centering
\includegraphics[width=0.3\columnwidth]{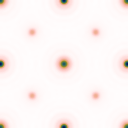}\hspace{3pt}        \includegraphics[width=0.3\columnwidth]{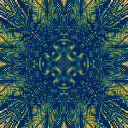}\hspace{3pt}        \includegraphics[width=0.3\columnwidth]{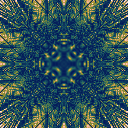}\hspace{3pt}\\[6pt]
\includegraphics[width=0.3\columnwidth]{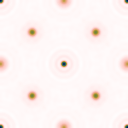}\hspace{3pt}        \includegraphics[width=0.3\columnwidth]{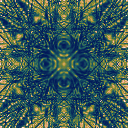}\hspace{3pt}        \includegraphics[width=0.3\columnwidth]{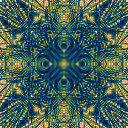}\hspace{3pt}\\[6pt]
\includegraphics[width=0.3\columnwidth]{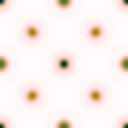}\hspace{3pt}        \includegraphics[width=0.3\columnwidth]{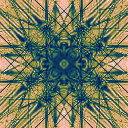}\hspace{3pt}        \includegraphics[width=0.3\columnwidth]{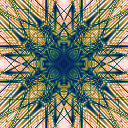}\hspace{3pt}\\[6pt]
\includegraphics[width=0.3\columnwidth]{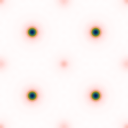}\hspace{3pt}        \includegraphics[width=0.3\columnwidth]{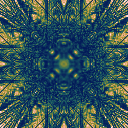}\hspace{3pt}        \includegraphics[width=0.3\columnwidth]{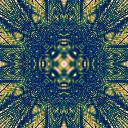}\hspace{3pt}\\[6pt]
%
\caption{Comparison of cGAN (central column) and \textsc{Felix} (right column) LACBED simulations for unseen $F\bar{4}3m$ crystal structures with identical projected atom locations (left column) at a thickness of $200$ nm. They are, with correlation $R$ top to bottom: half-Heusler ZrNi$_{1.056}$Sn ($R=0.919$, ICSD 194751), sphalerite InSb ($R=0.998$, ICSD 162196), sphalerite Cu$_{1.18}$Ge$_{3.63}$P$_{3.19}$ ($R=0.940$, ICSD 166923) and half-Heusler Fe$_{0.9}$Cu$_{0.03}$Sb ($R=0.895$, ICSD 152795). Colours match Fig.\ \ref{fig:projpot-ML-Felix}.}
\label{fig:pred-lacbed}
\end{figure}

\begin{figure*}[tb]
    \centering
    Intensity\\
    \includegraphics[scale=1]{figs/batloww_r_colorbar.pdf}\\[6pt]
    
    \makebox[0.3\columnwidth]{(a) Inputs}\hspace{3pt}   
    \makebox[0.3\columnwidth]{(b) 50 nm}\hspace{3pt}   
    \makebox[0.3\columnwidth]{(c) 100 nm}\hspace{3pt}   
    \makebox[0.3\columnwidth]{(d) 150 nm}\hspace{3pt}   
    \makebox[0.3\columnwidth]{(e) 200 nm}\\[6pt]

    \raisebox{0.15\columnwidth}{
  \hspace*{-0.18\columnwidth}{\makebox[0.16\columnwidth][l]{(A)\textsc{ Felix}}}}
    \includegraphics[width=0.3\columnwidth]{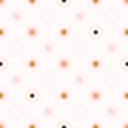}\hspace{3pt}   
    \includegraphics[width=0.3\columnwidth]{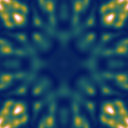}\hspace{3pt}   
    \includegraphics[width=0.3\columnwidth]{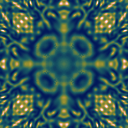}\hspace{3pt}   
    \includegraphics[width=0.3\columnwidth]{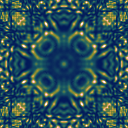}\hspace{3pt}   
    \includegraphics[width=0.3\columnwidth]{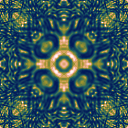}\hspace{0.2\columnwidth}
    \\[6pt]
    
    \raisebox{0.15\columnwidth}{
  \hspace*{-0.18\columnwidth}{\makebox[0.16\columnwidth][l]{(B) ML}}}
    \includegraphics[width=0.3\columnwidth]{assets/249064/potential.png}\hspace{3pt}   
    \includegraphics[width=0.3\columnwidth]{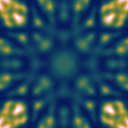}\hspace{3pt}   
    \includegraphics[width=0.3\columnwidth]{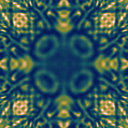}\hspace{3pt}   
    \includegraphics[width=0.3\columnwidth]{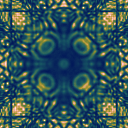}\hspace{3pt}   
    \includegraphics[width=0.3\columnwidth]{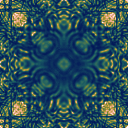}\hspace{0.2\columnwidth}
    \\[6pt]

    \raisebox{0.15\columnwidth}{
  \hspace*{-0.18\columnwidth}{\makebox[0.16\columnwidth][l]{(C) $p(i,j)$
  }}}\hspace{0.3\columnwidth}   \hspace{3pt}
    \includegraphics[width=0.3\columnwidth]{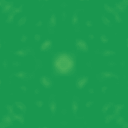}\hspace{3pt}   
    \includegraphics[width=0.3\columnwidth]{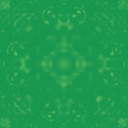}\hspace{3pt}   
    \includegraphics[width=0.3\columnwidth]{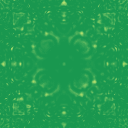}\hspace{3pt}   
    \includegraphics[width=0.3\columnwidth]{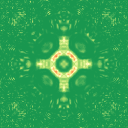}\hspace{0.2\columnwidth}
    \\[6pt]

    \centering
    Difference \\
    \includegraphics[scale=1]{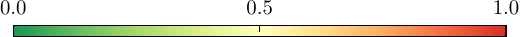}\\[6pt]
    \caption{ 
    Differences in {\sc Felix} (row A) and cGAN (row B)  bright field LACBED simulations for Dy$_2$Ti$_{1.9}$O$_{6.95}$ (ICSD {249064}, Fd$\overline{3}$mZ), an example close to the mean performance for each thickness. 
    Row C shows the pixel-by-pixel squared intensity differences \eqref{eq:pse}; cross-correlations $R$ are (a) $0.967$, (b) $0.926$, (c) $0.903$, and (d) $0.766$.}
    \label{fig:pred-lacbed-thickness}
\end{figure*}

A second trend in the quality of fit is seen in the inset of Fig.~\ref{fig:losses} standard errors for the complete dataset become slightly poorer as specimen thickness increases.  Examples of the discrepancies between the two methods at a specimen thickness of 200~nm are shown in Fig.~\ref{fig:pred-lacbed}. Here, it is apparent that very different LACBED patterns can be produced from structures with the same space group, ($F\bar{4}3m$) and even with atoms at the same coordinates in the projected unit cell.  Most of the differences between the cGAN and Bloch wave simulations are concentrated at the centre of the diffraction pattern, while details away from the centre are usually replicated very well.  This was a general trend across the data and is illustrated in Fig.~\ref{fig:pred-lacbed-thickness}.

In summary, these trends show that further improvements to the cGAN simulation are possible.


%

\begin{figure*}[tb]
    \centering
    Intensity\\
    \includegraphics[scale=1]{figs/batloww_r_colorbar.pdf}\\[6pt]

\makebox[0.3\columnwidth]{(a) 50 nm}\hspace{3pt}   
\makebox[0.3\columnwidth]{(b) 100 nm}\hspace{3pt}   
\makebox[0.3\columnwidth]{(c) 150 nm}\hspace{3pt}   
\makebox[0.3\columnwidth]{(d) 200 nm}\hspace{3pt}
\makebox[0.3\columnwidth]{}\\[6pt]

\raisebox{0.15\columnwidth}{
\hspace*{-0.10\columnwidth}{\makebox[0.18\columnwidth][l]{(A) Inputs}}}
\includegraphics[width=0.3\columnwidth]{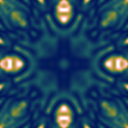}\hspace{3pt}   
\includegraphics[width=0.3\columnwidth]{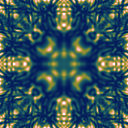}\hspace{3pt}        \includegraphics[width=0.3\columnwidth]{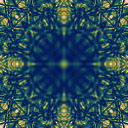}\hspace{3pt}        \includegraphics[width=0.3\columnwidth]{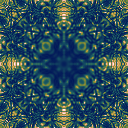}\hspace{3pt}\hspace{0.1\columnwidth}
\makebox[0.3\columnwidth][c]{(e) Ground truth}\\[6pt]

\raisebox{0.15\columnwidth}{
\hspace*{-0.10\columnwidth}{\makebox[0.18\columnwidth][l]{(B) ML}}}
\includegraphics[width=0.3\columnwidth]{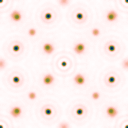}\hspace{3pt}   
\includegraphics[width=0.3\columnwidth]{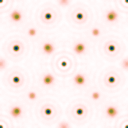}\hspace{3pt}        \includegraphics[width=0.3\columnwidth]{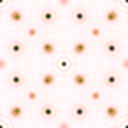}\hspace{3pt}        \includegraphics[width=0.3\columnwidth]{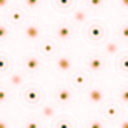}\hspace{0.1\columnwidth}\hspace{3pt}
\includegraphics[width=0.3\columnwidth]{assets/196220/ml_pattern_0500.png}\hspace{3pt}
\\[6pt]

\raisebox{0.15\columnwidth}{
\hspace*{-0.10\columnwidth}{\makebox[0.18\columnwidth][l]{(C) $p(i,j)$}}}
\includegraphics[width=0.3\columnwidth]{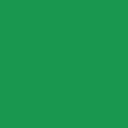}\hspace{3pt}   
\includegraphics[width=0.3\columnwidth]{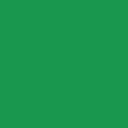}\hspace{3pt}        \includegraphics[width=0.3\columnwidth]{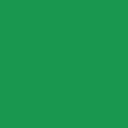}\hspace{3pt}        \includegraphics[width=0.3\columnwidth]{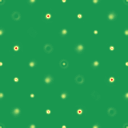}\hspace{6pt}\makebox[0.4\columnwidth][l]{}\\[6pt]
\centering
Difference \\
\includegraphics[scale=1]{figs/rainbow_continuous_colorbar.pdf}\\[6pt]
\caption{
    Thickness dependence of cGAN reconstructions of projected potential $\rho(\mathbf{r})$ from $[001]$ bright field LACBED patterns. This example is CrFeMg$_{0.14}$O$_4$Zn$_{0.86}$, $a=8.406\AA$ (ICSD 196220, Fd$\overline{3}$mS). 
    Columns (a) to (d) indicate specimen thickness, while (e) gives the ground truth projected potential $\rho(\mathbf{r})$.
    Row (A) gives the normalized LACBED intensities (see top colour bar) from {\sc Felix} and row (B) the cGAN-generated projected potentials. 
    Row (C) shows pixel-by-pixel differences following Eq.\ \eqref{eq:pse} with differences indicated by the bottom colour bar. 
    The cross-correlation $R$ between {\sc Felix}-simulated LACBEDs and ML-predicted $\rho(\mathbf{r})$ potentials for this structure are a) $0.9996$, b) $0.9992$, c) $0.9995$, and d) $0.6326$. 
\label{fig:pred-projpot-thickness}
}
\end{figure*}

\subsection{Predicting the Projected potential}
\label{sec:electron_potential}

We now turn to the inverse problem, i.e., producing a real space projected potential, given the $[001]$ bright field LACBED patterns as input. There is no conventional method which can perform this calculation.  In Fig.~\ref{fig:pred-projpot-thickness}, we show an example \emph{inverse} calculation using the same cGAN structure as detailed in Fig.\ \ref{fig:cGAN-algo-structure}.  As before, the training was over ten independent learning cycles. We find that this inverse generative approach works, on average, as well as the original workflow from projected potential Eq.~\eqref{eq:projpot} to LACBED pattern, with mean $\ell_\text{MSE}$ of $2.5$, $2.3$, $2.4$ and $2.4$ ($\times 10^{-4}$) for thicknesses $50$, $100$, $150$ and $200$ nm, respectively.  The data is positively skewed, and for more than half of our predictions, $\ell_\text{MSE} < 1\times 10^{-4}$.  This indicates that there is an excellent calculation of projected potential for the majority of input LACBED patterns, with a minority of results that give a poor fit index.

Examination of examples with relatively high $\ell_\text{MSE}$ show that the apparently poor performance is related once more to the origin of the unit cell.  In Fig.~\ref{fig:pred-projpot-thickness} the difference between the cGAN-calculated potential -- row (B) -- and ground truth -- Fig.~\ref{fig:pred-projpot-thickness}(e) -- is shown in row (C).  This is very low for thicknesses 50-150 nm, columns (a)-(c), but large for column (d), 200 nm.  On closer inspection of the calculated potential in (d) it is apparent that it is essentially the same as the other calculations, but shifted by $1/8[110]$.  The chosen metric $\ell_\text{MSE}$ therefore underestimates the performance of the method and there is good reason to think that with a more appropriate metric, which is insensitive to origin choice, even better performance can readily be achieved.
%

%

\section{Discussion and conclusions}
\label{sec:discussion_and_conclusions}

%

Although the computing architectures are different (multi-CPU + MPI, vs GPU) \textsc{Felix} Bloch-wave calculations complete in $400$\,seconds while the cGAN reconstruction takes $2\times 10^{-2}$\,seconds. This is a considerable speedup. Similar improvements may be possible in comparison with multislice calculations.

This work should be seen as a proof of principle, and there remains considerable room for improving the approach as presented here.  The geometry of the calculation was chosen to match common cGAN inputs and outputs, i.e., $8$-bit images of moderate size with dimensions that are a power of $2$.  A crystal which is not cubic, or even a cubic one in an arbitrary orientation, does not conform to this geometry, nor does the electron diffraction pattern it would produce.  The use of a large-angle convergent electron probe is also relatively unusual.  Nevertheless, these are essentially details that must be dealt with for any particular implementation of the method, and there appears to be no fundamental barrier to exploiting the approach across the full range of scattering geometries commonly used in electron diffraction (including methods that extract data in the diffraction plane, like scanning transmission electron microscopy, STEM) \cite{Friedrich2023PhaseLearning}.  Even in our specific, very limited, implementation some improvements are obviously required, in particular reducing the sensitivity to the choice of coordinate origin of the unit cell in the forward calculation, and use of a similarly insensitive metric for the inverse calculation.  Similarly, we used a strictly defined angular range and only reconstructed bright field LACBED patterns, but different angular ranges and many other $h,k,l$ diffracted beams are readily available that could both be simulated and used in an inverse calculation.  Other (supervised) machine learning architectures may prove equally suitable to these applications, such as variational autoencoders \cite{Partington2022UsingPatterns}.

Perhaps the most exciting aspect of this work is the apparent ease with which the solution of the inverse problem is approximated by a machine learning approach.  If the initial promise shown here is fulfilled, development of this method could give a valuable new approach to the study of materials structure and quantum chemistry.
Nevertheless, it remains to be demonstrated that this method can be effective with experimental data and, in any implementation, a suitable training data set is necessary. Data augmentation, which has been proven to better generalise deep learning models on limited datasets \cite{Shorten2019}, may reduce this dependency.  Furthermore, the \textsc{pix2pix} architecture allows for the use of a semantic map, which we did not explore; future work could include (chemical) information on the $z$ coordinate of charges, which is lost in our projection.  

In summary, we have shown that a generative machine learning architecture trained on crystal structure data can reconstruct bright field LACBED patterns that are of comparable quality to those calculated by standard simulation packages, such as the Bloch-wave method.  Speed improvements are significant, roughly four orders of magnitude in our implementation.  A calculation of the projected potential of the crystal, using the diffraction pattern as input, appears to be equally rapid and accurate.

\section*{Acknowledgements}
RB and RAR thank James Partington and Jeremy Thornton for valuable discussions during the early stages of this work. JJW was financially supported by the Undergraduate Research Support Scheme (URSS) at the University of Warwick. At Oxford, JJW is grateful to Worcester College for a Master of Mathematical Sciences Scholarship and to the Mathematical Institute for providing a bursary. 
We are especially appreciative for the high-performance computing facilities and the Sulis Tier 2 HPC platform provided and hosted by the Scientific Computing Research Technology Platform (RTP). Sulis is funded by EPSRC Grant EP/T022108/1 and the HPC Midlands+ consortium.
UK research data statement: The code for data generation \cite{BeanlandFelix:Software}, the generated data \cite{Webb2023FelixPatterns} and all machine learning is published open source \cite{Webb2024GitHubWephy/ai-diffraction}.

\bibliographystyle{model1-num-names}
\bibliography{JJW.bib, Extra_references.bib}

\ifNOSUP\end{document}\else%

\clearpage\newpage
\setcounter{section}{0}
\setcounter{figure}{0}
\setcounter{table}{0}
\def\thesection{S\arabic{section}}
\def\thefigure{S\arabic{figure}}
\def\thetable{S\arabic{table}}
\setcounter{page}{1}
\pagestyle{plain}

{\center

\textbf{Supplemental Material}\\[2ex]

\textbf{Large-Angle Convergent-Beam Electron Diffraction Patterns via Conditional Generative Adversarial Networks}\\[2ex]

\noindent%
Joseph J Webb, Richard Beanland, Rudolf A R\"{o}mer
\\[2ex]

{Department of Physics, University of Warwick, Coventry, CV4 7AL, UK}\\

}
\hspace*{2ex}

\section{LACBED}

text text

\section{Projected potential}

text text 



\fi\end{document}